\documentclass[sn-mathphys]{sn-jnl}% Math and Physical Sciences Reference Style
\usepackage{libertine}
\usepackage{url}
\usepackage{bbm}
\usepackage{longtable}
\usepackage{times}
\usepackage{hyperref}
\usepackage{nicefrac}
\usepackage{amsfonts}
\usepackage{bm}
\usepackage{multirow}
\usepackage{mathrsfs}
\usepackage{revsymb}
\usepackage{amssymb}
\usepackage{amsmath}
\usepackage{mathtools}
\usepackage{slashed}
\usepackage{dcolumn}
\usepackage[babel]{csquotes}
\usepackage{lineno}
\usepackage{subfigure}
\usepackage{tabularx}
\jyear{2023}%
\begin{document}

\title[$Q$-value determination of the $^{163}\mathrm{\textit{Ho}}$ electron capture decay]{Penning-trap measurement of the $Q$-value of the electron capture in $^{163}\mathrm{Ho}$ for the determination of the electron neutrino mass}

\author*[1]{\fnm{Christoph} \sur{Schweiger}}\email{christoph.schweiger@mpi-hd.mpg.de}
\author[2]{\fnm{Martin} \sur{Bra{\ss}}}
\author[1]{\fnm{Vincent} \sur{Debierre}}
\author[1]{\fnm{Menno} \sur{Door}}
\author[3]{\fnm{Holger} \sur{Dorrer}}
\author[3,4,5]{\fnm{Christoph E.} \sur{Düllmann}}
\author[6]{\fnm{Christian} \sur{Enss}}
\author[1]{\fnm{Pavel} \sur{Filianin}}
\author[6]{\fnm{Loredana} \sur{Gastaldo}}
\author[1]{\fnm{Zolt\'an } \sur{Harman}}
\author[2]{\fnm{Maurits W.} \sur{Haverkort}}
\author[1]{\fnm{Jost} \sur{Herkenhoff}}
\author[7]{\fnm{Paul} \sur{Indelicato}}
\author[1]{\fnm{Christoph H.} \sur{Keitel}}
\author[1]{\fnm{Kathrin} \sur{Kromer}}
\author[1,8]{\fnm{Daniel} \sur{Lange}}
\author[9]{\fnm{Yuri N.} \sur{Novikov}}
\author[3,4]{\fnm{Dennis} \sur{Renisch}}
\author[1]{\fnm{Alexander} \sur{Rischka}}
\author[1]{\fnm{Rima X.} \sur{Schüssler}} 
\author[1]{\fnm{Sergey} \sur{Eliseev}}
\author[1]{\fnm{Klaus} \sur{Blaum}}

\affil[1]{\orgname{Max-Planck-Institut für Kernphysik}, \orgaddress{\street{Saupfercheckweg 1}, \city{Heidelberg}, \postcode{69117}, \country{Germany}}}
\affil[2]{\orgdiv{Institute for Theoretical Physics}, \orgname{Heidelberg University}, \orgaddress{\street{Philosophenweg 19}, \city{Heidelberg}, \postcode{69120}, \country{Germany}}}
\affil[3]{\orgdiv{Department Chemie - Kernchemie}, \orgname{Johannes Gutenberg-Universität Mainz}, \orgaddress{\street{Fritz-Straßmann-Weg 2}, \city{Mainz}, \postcode{55128}, \country{Germany}}}
\affil[4]{\orgname{Helmholtz-Institut Mainz}, \orgaddress{\street{Staudingerweg 18}, \city{Mainz}, \postcode{55128}, \country{Germany}}}
\affil[5]{\orgname{GSI Helmholtzzentrum für Schwerionenforschung GmbH}, \orgaddress{\street{Planckstraße 1}, \city{Darmstadt}, \postcode{64291}, \country{Germany}}}
\affil[6]{\orgdiv{Kirchhoff-Institute for Physics}, \orgname{Heidelberg University}, \orgaddress{\street{Im Neuenheimer Feld 227}, \city{Heidelberg}, \postcode{69120}, \country{Germany}}}
\affil[7]{\orgdiv{Laboratoire Kastler Brossel, CNRS, ENS-PSL Research University, Collège de France, Campus Pierre et Marie Curie}, \orgname{Sorbonne Université}, \orgaddress{\street{4 place Jussieu}, \city{Paris}, \postcode{75005}, \country{France}}}
\affil[8]{\orgname{Heidelberg University}, \orgaddress{\street{Grabengasse 1}, \city{Heidelberg}, \postcode{69117}, \country{Germany}}}
\affil[9]{\orgname{NRC “Kurchatov Institute”-Petersburg Nuclear Physics Institute}, \orgaddress{\city{Gatchina}, \postcode{188300}, \country{Russia}}}
%\affil[10]{\orgname{Van der Waals–Zeeman Institute, Institute of Physics, University of Amsterdam}, \orgaddress{\street{Science Park 904}, \city{Amsterdam}, \postcode{1098XH}, \country{The Netherlands}}}

\presentaddress{
R.X. Schüssler: Van der Waals–Zeeman Institute, Institute of Physics, University of Amsterdam, Science Park 904, Amsterdam 1098XH, The Netherlands\\
Martin Bra{\ss}: Institute of Solid State Physics, TU Wien, 1040 Vienna, Austria}

\abstract{The investigation of the absolute scale of the effective neutrino mass remains challenging due to the exclusively weak interaction of neutrinos with all known particles in the standard model of particle physics.
Currently, the most precise and least model-dependent upper limit on the electron antineutrino mass is set by the KATRIN experiment from the analysis of the tritium $\beta$-decay. 
Another promising approach is the electron capture in $^{163}\mathrm{Ho}$, which is under investigation using microcalorimetry within the ECHo and HOLMES collaborations.
An independently measured $Q$-value of this process is vital for the assessment of systematic uncertainties in the neutrino mass determination.

Here, we report a direct, independent determination of this $Q$-value by measuring the free-space cyclotron frequency ratio of highly charged ions of $^{163}\mathrm{Ho}$ and $^{163}\mathrm{Dy}$ in the Penning trap experiment \textsc{Pentatrap}.
Combining this ratio with atomic physics calculations of the electronic binding energies yields a $Q$-value of $2863.2(0.6)\,\mathrm{eV}/c^{2}$ - a more than 50-fold improvement over the state-of-the-art. 
This will enable the determination of the electron neutrino mass on a sub-eV level from the analysis of the electron capture in $^{163}\mathrm{Ho}$.}


%%\pacs[JEL Classification]{D8, H51}

%%\pacs[MSC Classification]{35A01, 65L10, 65L12, 65L20, 65L70}

\maketitle

\section{The absolute scale of the neutrino mass}\label{sec:intro}

The observation of the neutrino flavor oscillations proves that neutrinos are massive particles, establishing that the weak neutrino flavor eigenstates are a superposition of three neutrino-mass eigenstates in contradiction to the Standard Model of particle physics~\cite{Fukuda98, Ahmad02}.
In oscillation experiments merely the differences of the squared neutrino mass eigenvalues can be investigated, leaving the absolute scale of the neutrino mass an open question.
Thus, the absolute scale of the neutrino mass remains one of the most sought-after quantities in nuclear and particle physics, cosmology and beyond Standard Model theories that could potentially explain the origin of the neutrino rest mass~\cite{King03, Drexlin13, Formaggio21, Gouvea16}.

Neutrinos are produced in weak nuclear decays; a model-independent measurement of their rest mass can be performed in a kinematic study of the decay products, where the neutrino itself is not directly detected. 
Relying on energy and momentum conservation, this is currently the most model-independent approach for neutrino mass determinations.
Kinematic investigations constrain the effective rest mass of the electron neutrino or antineutrino \mbox{$m_{\nu_{e}}^{2} = \sum_{i=1}^{3} \lvert U_{ei}\rvert^{2} m_{i}^{2}$}, where $U_{fi}$ are the elements of the Pontecorvo–Maki–Nakagawa–Sakata (PMNS) matrix, which describes the superposition of mass eigenstates $m_{i}$ (\mbox{$i\in\{1,2,3\}$}) in the flavor eigenstates $\nu_{f}$ (\mbox{$f\in\{e,\mu,\tau\}$}).
The individual mass eigenstates are not resolved in these experiments since the squared mass differences are well below current instrumental resolutions, with the largest one being~$\Delta m_{32}^{2} = (2.453 \pm 0.033) \cdot 10^{-3}\,\mathrm{eV}^{2}/c^{4}$~\cite{pdg22}.

The most stringent constraint on the neutrino mass scale comes from the analysis of the matter distribution in the universe which results in a limit on the sum of the neutrino masses of~$<120\,\mathrm{meV}/c^{2}$~\cite{Aghanim20} while the most stringent direct limit of~$0.8\,\mathrm{eV}/c^{2}$ (90 \% C.L.) from a kinematic study of the tritium $\beta$-decay  is set by the KATRIN collaboration~\cite{Aker21, Aker22}.

Complementary to this approach, there are several experiments using calorimetric techniques to investigate the neutrino rest mass directly. 
Historically, the first calorimetric approaches were the MANU and MIBETA experiments investigating the $^{187}\mathrm{Re}$ $\beta$-decay yielding upper limits of $19$ and $15\,\mathrm{eV}^{2}/c^{2}$ (90\% C.L.), respectively\cite{Nucciotti16}.
Two current experiments, namely ECHo~\cite{Gastaldo17, Velte19} and HOLMES~\cite{Faverzani16, Nucciotti18}, investigate the electron-capture in $^{163}\mathrm{Ho} \,\rightarrow\, ^{163}\mathrm{Dy} + \nu_{e} + E_{\mathrm{cal}}$, with $E_{\mathrm{cal}}$ being the energy detected in a calorimeter. 
The current upper limit of the electron neutrino rest mass is on a level of $<150\,\mathrm{eV}/\mathrm{c}^{2}$~\cite{Velte19} and the ECHo and HOLMES collaborations aim to achieve sensitivities well below $<1\,\mathrm{eV}/\mathrm{c}^{2}$~\cite{Gastaldo17}.

Within the ECHo collaboration, metallic magnetic calorimeters are used for the measurement of the energy of all emitted radiation except for the energy carried away by the neutrino. This is obtained by implanting $^{163}\mathrm{Ho}$ ions directly into the absorber material of the detector.
The calorimetrically measured decay spectrum is subsequently analyzed by fitting it to a theoretical spectral shape from which the $Q$-value as well as the effective electron neutrino mass $m_{\nu_{e}}$ can be determined.
In order to quantitatively investigate systematic effects in the interpretation of the calorimetrically measured spectra, that might arise due to the $^{163}\mathrm{Ho}$ ions being implanted into a metallic material, this $Q$-value is best compared to one obtained from an independent direct measurement.
The required accuracy of $\sim 1\,\mathrm{eV}/c^2$ can currently only be reached using high-precision Penning-trap mass spectrometry (PTMS).
In PTMS, the $Q$-value is addressed directly through a measurement of the mass difference of the mother and daughter nuclides, $^{163}\mathrm{Ho}$ and $^{163}\mathrm{Dy}$, respectively~\cite{Gastaldo17, Eliseev13a}, by measuring the free cyclotron frequency ratio of the two species in a strong homogeneous magnetic field $B$.
Within a magnetic field, an ion with charge-to-mass ratio $q/m$ is forced onto a circular orbit where it revolves with the free-space cyclotron frequency $\nu_{c} = \frac{1}{2 \pi} \frac{q}{m} B$.
In a Penning trap, a superimposed weak quadrupolar electrostatic potential confines the ion along the magnetic field lines and modifies the ion's radial motion:
The free-space cyclotron motion splits into the magnetron motion with the frequency $\nu_{-}$ and the modified cyclotron motion with frequency $\nu_{+}$. 
In addition, the quadrupolar electrostatic potential induces a harmonic oscillatory motion with frequency $\nu_{z}$ along the magnetic field lines.
From a measurement of all three motional eigenfrequencies, the free-space cyclotron frequency can be reconstructed using the invariance theorem $\nu_{c}^{2} = \nu_{+}^{2} + \nu_{z}^{2} + \nu_{-}^{2}$~\cite{Brown86}.
From subsequent measurements of the free-space cyclotron frequency the ratio $R_{q+} = \nu_{c}(^{163}\mathrm{Dy}^{q+})/\nu_{c}(^{163}\mathrm{Ho}^{q+})$ is determined, which finally allows the determination of the $Q$-value by including atomic physics calculations of the binding energy difference $\Delta E_{\mathrm{B}}^{q+}$ of the removed electrons:
\begin{equation}
Q = m_{\mathrm{Dy}}^{q+} \left(R_{q+}-1\right) + \Delta E_{\mathrm{B}}^{q+}.
\label{eq:Qvaluecalculation}
\end{equation}
$\Delta E_{\mathrm{B}}^{q+}$ is given by the difference in the sum of the binding energies of the $n$ missing electrons in the highly charged ions (HCIs) of both nuclides and $m_{\mathrm{Dy}}^{q+}$ is the ``reference'' mass of the HCI of dysprosium.
$q = n \cdot e$ is the charge of the ions, with $e$ being the elementary charge and $n$ the number of removed electrons (``charge state''). In order to enhance the readability in formulas, sometimes $q$ also denotes the number of missing electrons $n$.

\section{The Penning-trap experiment \textsc{Pentatrap}}\label{sec:PTMS}

\begin{figure}[htbp!]
	\centering
		\includegraphics[width=0.60\textwidth]{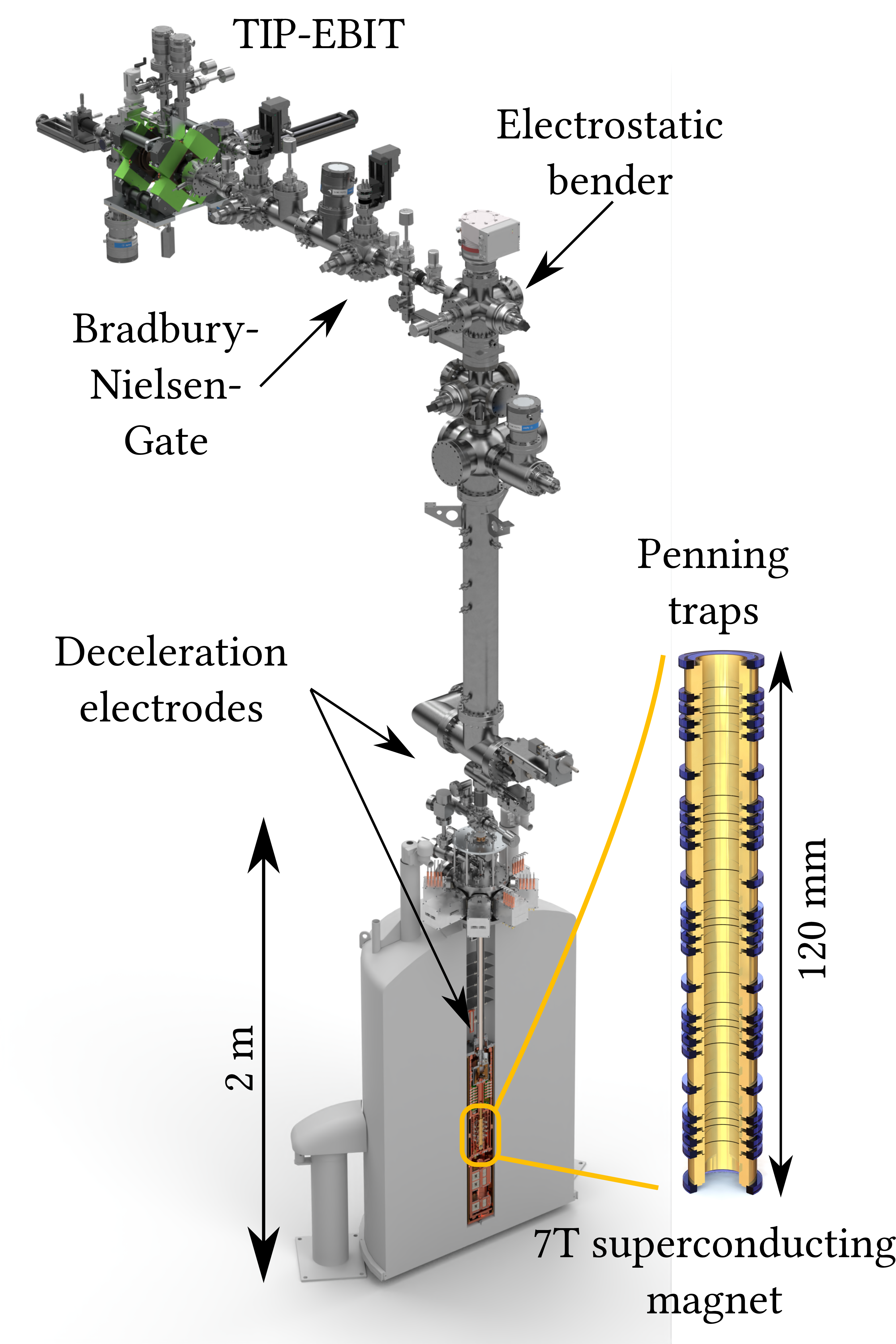}
	\caption{Rendered overview of the \textsc{Pentatrap} experimental setup.
	The upper horizontal part of the beamline is located on the ground floor while the superconducting magnet is located in a dedicated laboratory in the basement.
	The TIP-EBIT is an electron beam ion trap specifically designed for very small samples sizes~\cite{Schweiger19}.
	Follwing the TIP-EBIT in the horizontal beamline, a Bradbury-Nielsen gate is used to separate a single charge state.
	HCIs produced in the TIP-EBIT are guided through the electrostatic beamline to the stack of five identical Penning traps in the superconducting magnet.
	For capturing the HCIs in the Penning traps deceleration electrodes with appropriately timed voltage pulses are used.
	A more detailed view of the Penning trap stack is shown on the right.
	}
	\label{fig:PentatrapSetup}
\end{figure}

\paragraph{Experimental setup} 
The measurement of the free-space cyclotron frequency ratio $R$ of the two HCIs $^{163}\mathrm{Ho}^{q+}$ and $^{163}\mathrm{Dy}^{q+}$ was carried out with the high-precision Penning-trap mass spectrometer \textsc{Pentatrap} located at the Max-Planck-Institute for Nuclear Physics in Heidelberg, Germany \cite{Repp12, Filianin21}. 
An overview of the apparatus is given in Figure~\ref{fig:PentatrapSetup}.

HCIs of the synthetic radioisotope $^{163}\mathrm{Ho}$, which was produced by neutron irradiation of stable $^{162}\mathrm{Er}$~\cite{Dorrer18}, and HCIs of the stable $^{163}\mathrm{Dy}$ are produced in a compact room-temperature electron beam ion trap (EBIT) that is specifically designed and constructed for the production of HCIs from samples available only in limited quantities (TIP-EBIT)~\cite{Schweiger19}.
For the measurements reported here only $2 \cdot 10^{15}$ atoms of $^{163}\mathrm{Ho}$ were used, with a typical sample containing about $10^{14}$ atoms of $^{163}\mathrm{Ho}$.
HCIs of the two species are extracted with a kinetic energy of $4.4\,\mathrm{keV}/q$ from the EBIT and transported through an electrostatic beamline towards the Penning traps.
Individual charge states $n = \{38,\,39,\,40\}$ are selected using a Bradbury-Nielsen Gate and a fast switching electronic circuit~\cite{Bradbury36, Schweiger22} located about $1.5\,\mathrm{m}$ from the EBIT.
Just before reaching the mass spectrometer, the HCIs are decelerated to a few $\mathrm{eV}/q$ by appropriately timed voltage pulses on two cylindrical drift tubes.

\begin{figure}[htbp!]
	\centering
		\includegraphics[width=\textwidth]{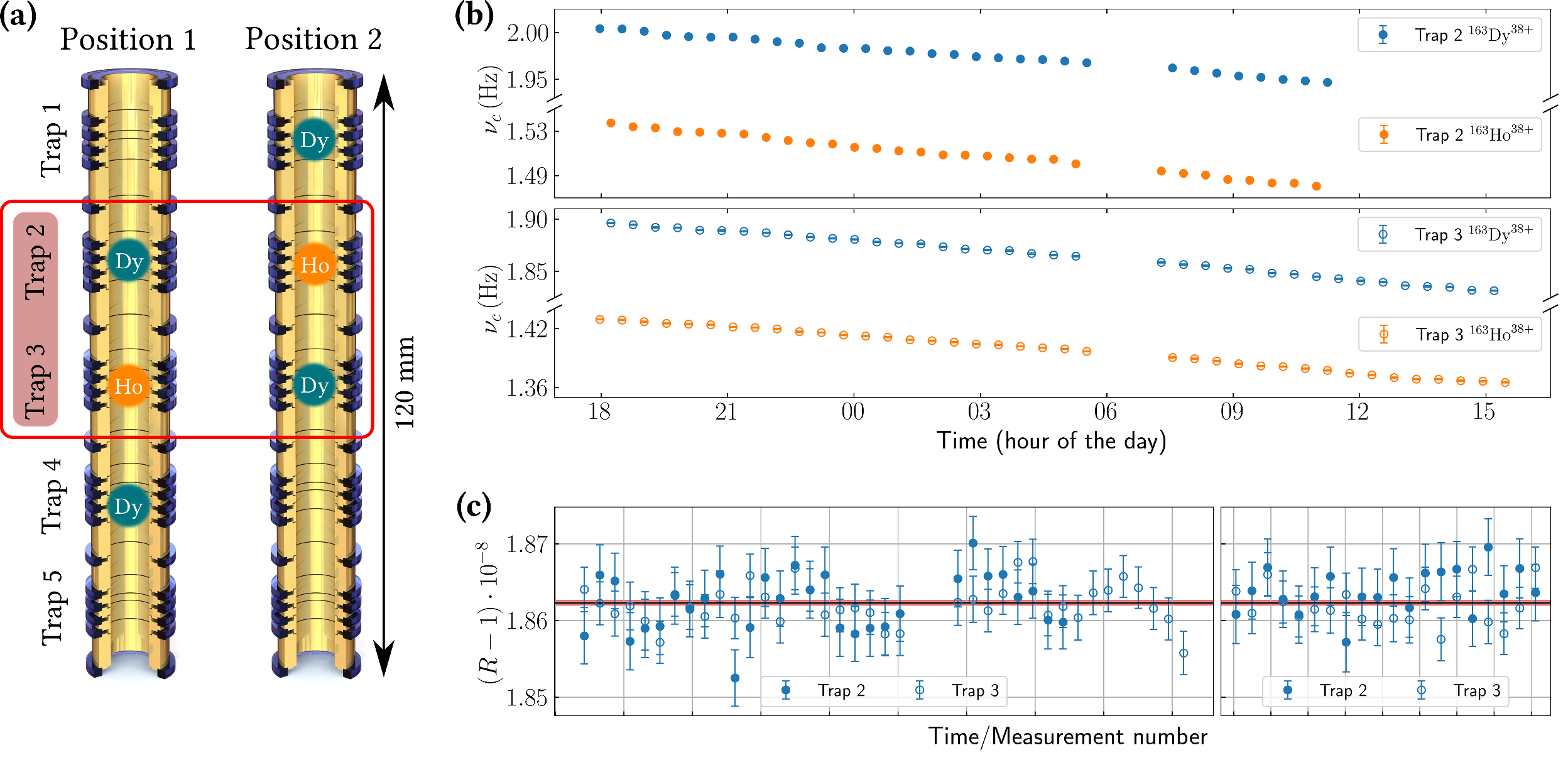}
	\caption{Overview of the measurement procedure and resulting data structure.
	Error bars correspond to the $1\sigma$ statistical uncertainty which is propagated by Gaussian uncertainty propagation.
	(a) Rendering of the stack of five identical, cylindrical Penning traps of the \textsc{Pentatrap} experiment. 
	Traps 2 and 3, with labels marked in red, are used as measurement traps and are equipped with a detection system.
	Shuttling the ions from Position 1 into Position 2 effectively swaps the ion species in traps 2 and 3 resulting in the alternating data structure as shown in panel (b).
The traps 1 and 4 are used as storage traps while trap 5 is not used in this measurement.
	(b) Exemplary dataset of the measured free cyclotron frequencies $\nu_{c}$ of $^{163}\mathrm{Ho}$ (orange) and $^{163}\mathrm{Dy}$ (blue) in the traps 2 (upper panel) and 3 (lower panel) for one measurement run in charge state $q = 38\cdot e$.
	For trap 2 and 3 frequency offsets of $25081589\,\mathrm{Hz}$ and $25081620\,\mathrm{Hz}$ were subtracted.
	The linear drift of the free cyclotron frequency which can be attributed to the slow decay of the magnetic field of the superconducting magnet due to the flux creep effect~\cite{Anderson62, Anderson64}.
	Please note that the vertical axis is broken for illustrative purposes while there are no left and right sub-panels.
	(c) Ratios $R_{i}$ of the free cyclotron frequencies $\nu_{c}$ of $^{163}\mathrm{Dy}$ and $^{163}\mathrm{Ho}$ in traps 2 (filled circles) and 3 (empty circles) determined from the full dataset of two runs for the charge state $n=38$.
	The data of each run is shown in a dedicated sub-panel where the ratios from (b) are shown in the left sub-panel.
	The horizontal black line indicates the weighted average of all measured ratios for this charge state with the light red band marking the 1$\sigma$ uncertainty band.}
	\label{fig:ExampleData}
\end{figure}

The mass spectrometer consists of a stack of five identical, cylindrical Penning traps located in the cold bore of a~$7\,\mathrm{T}$, actively shielded superconducting magnet~\cite{Roux12, Repp12}.
The voltages applied to the Penning-trap electrodes are supplied from an ultra-stable voltage source~\cite{Boehm16}.
The Penning traps as well as the detection system are located inside a vacuum chamber immersed in liquid helium at a temperature of about $4\,\mathrm{K}$.
Two (trap~2 and trap~3, cf. Fig.~\ref{fig:ExampleData}~(a)) of the five Penning traps are equipped with a non-destructive image-current detection system~\cite{Wineland75, Feng96, Repp12, Nagahama16} and are used for the measurement of the ions' motional frequencies.
Trap 1 and trap 4 serve as storage traps while trap 1 is also used as a capture trap when a new set of ions is loaded into the trap stack.

Environmental parameters affecting the magnetic field in the traps are stabilized e.g. the temperature in the laboratory to $0.1\,\mathrm{K} / \mathrm{day}$ as well as the liquid helium level and pressure of helium gas inside the cold bore of the magnet.
In these conditions the magnetic field exhibits a relative drift of a few $10^{-10}$ per hour~\cite{Kromer22}. 
Frequency measurements are performed overnight and on weekends when external perturbations are minimal.

The measurement starts with loading a set of three ions, in the order $^{163}\mathrm{Dy}$, $^{163}\mathrm{Ho}$ and $^{163}\mathrm{Dy}$ into traps~2,~3, and~4, respectively (cf.~Fig.~\ref{fig:ExampleData}~(a)).
The motional frequencies of the HCIs in traps 2 and 3 are measured simultaneously, starting with the ions in position 1.
Subsequently, the ions are shuttled to position 2, which effectively swaps the ion species in traps 2 and 3 (cf.~Fig.~\ref{fig:ExampleData}~(a)) and the measurement is repeated.
The resulting data structure is shown in Fig.~\ref{fig:ExampleData}~(b) where the free-space cyclotron frequency $\nu_{c}$ is plotted as a function of the measurement time. 
Alternating datapoints for $^{163}\mathrm{Dy}$ and $^{163}\mathrm{Ho}$ result from the swapping of the ion species in traps 2 and 3. More details on the ion preparation and the measurement sequence is given in the Methods Section~\ref{sec:methods}.

\paragraph{Data analysis}
In order to extract frequency ratios $R$ from the free-space cyclotron frequencies $\nu_{c}$, the magnetic field behavior has to be interpolated in-between the individual frequency measurement datapoints from one species to the time when the other species' frequencies were measured.

Fig.~\ref{fig:ExampleData}~(b) shows exemplary the free cyclotron frequencies from one measurement run performed on ions with the charge state $q=38\cdot e$.
The linear slope of the data points can be attributed to the slow decay of the magnetic field of the superconducting magnet due to the flux creep effect~\cite{Anderson62, Anderson64} and is on the order of a few $10^{-10}$ per hour relative to the absolute magnetic field of $\sim 7\,\mathrm{T}$.

\begin{figure}[htbp!]
	\subfigure[]{
	\includegraphics[width=0.49\textwidth]{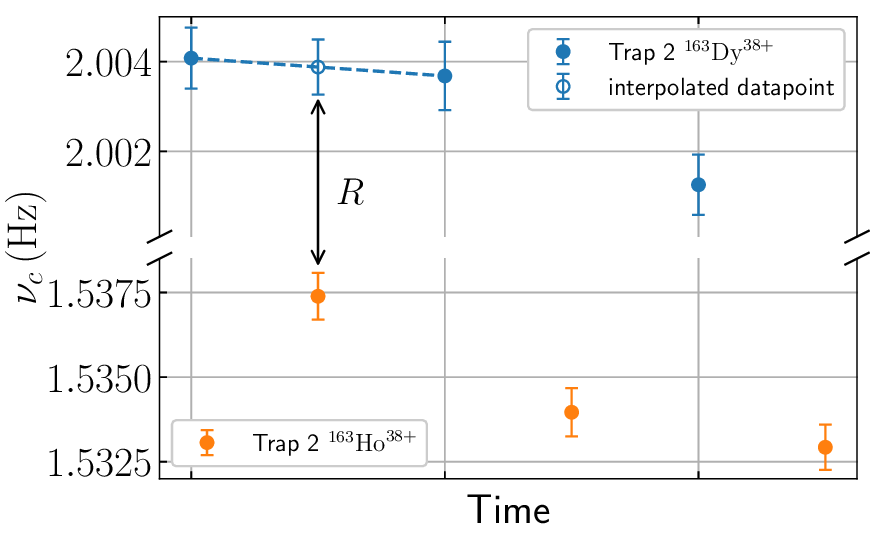}
	}
	\subfigure[]{
	\includegraphics[width=0.49\textwidth]{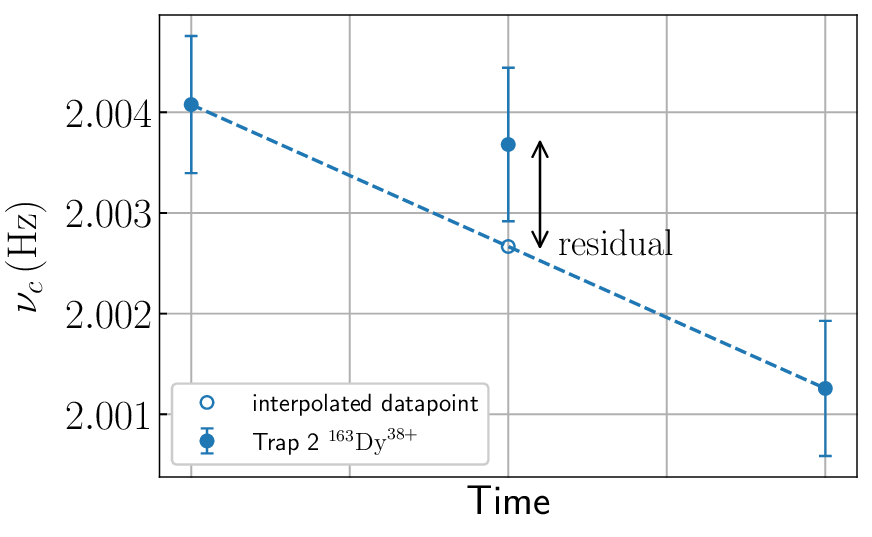}
	}
\caption{Detailed plot of the first few datapoints of the cyclotron frequency $\nu_{c}$ from Figure~\ref{fig:ExampleData} (b) in order to illustrate the data analysis procedure. 
From the frequency values an offset of $25081589\,\mathrm{Hz}$ is subtracted.
For details on the analysis procedure see main text.
Error bars correspond to the $1\sigma$ statistical uncertainty which is propagated by Gaussian uncertainty propagation.
(a) Linear interpolation between two $^{163}\mathrm{Dy}$ datapoints to the time at which $^{163}\mathrm{Ho}$ was measured for the determination of the free-space cyclotron frequency ratio $R$.
Please note that the vertical axis was broken for illustration purposes.
(b) Exemplary estimation of the non-linearity by interpolation of the data onto itself.
Here we linearly interpolate between the first and third datapoint and determine the difference between the measured datapoint in between and the interpolated one. 
The sum of these ``residuals'' divided by the number of residuals in the full dataset is taken into account as an additional uncertainty on the ratio.}
\label{fig:DataAnalysis}
\end{figure}

In the data analysis, the frequency of $^{163}\mathrm{Dy}$ is linearly interpolated between two datapoints to the time where $^{163}\mathrm{Ho}$ was measured.
From this interpolated datapoint the ratio $R$ is determined as illustrated in Figure~\ref{fig:DataAnalysis}~(a).
This procedure is followed for the full dataset.
Residual non-linear behavior of the cyclotron frequency drift, originating from physical effects that alter the temperature and position of magnetic materials that surround the Penning traps and change the magnetic field within the traps is taken into account in the uncertainty of the interpolated R.
For this, the frequency data points are interpolated back to themselves (see Figure~\ref{fig:DataAnalysis}~(b)), and the sum of the residuals divided by the number of residuals is included as an additional uncertainty in the ratio.
The resulting ratios $R_{i} = \nu_{c,i}(^{163}\mathrm{Dy}^{38+})/\nu_{c,i}(^{163}\mathrm{Ho}^{38+})$ for the two measurement runs are shown in Fig.~\ref{fig:ExampleData}~(c) for both traps. 
The ratios for the individual traps are consistent, therefore the final ratio is calculated as the weighted average and shown as a red line including the $1\sigma$ uncertainty band. 
For the calculation of the uncertainty of the final ratio, the inner error $\sigma_{\mathrm{int}}^{2}$ and the outer error $\sigma_{\mathrm{ext}}^{2}$ are calculated and the larger of the two is used as the final uncertainty~\cite{Nagy08, Birge32}:

\begin{align}
\sigma_{\mathrm{int}}^{2} &= \frac{1}{\sum_{i}\frac{1}{\sigma_{i}^{2}}} \\
\sigma_{\mathrm{ext}}^{2} &= \frac{\sum_{i}\frac{1}{\sigma_{i}^{2}} (R_{i}-\tilde{R})^{2}}{(N-1) \sum_{i}\frac{1}{\sigma_{i}^{2}}}.
\end{align}
Here, $R_{i}$ and $\sigma_{i}$ are individual cyclotron frequency ratios and their corresponding $1\sigma$ uncertainty, $\tilde{R}$ is the weighted average and $N$ is the total number of ratios.

The total systematic uncertainty (e.g. field anharmonicities and inhomogeneity, image charge shift and relativistic shift) is strongly suppressed due to the fact that $^{163}\mathrm{Ho}$ and $^{163}\mathrm{Dy}$ in the same charge state form a unique mass doublet with a sufficiently small mass difference of about $2.8\,\mathrm{keV}$.
With a difference in mass-to-charge ratio of only about $10^{-8}$ the same trapping potential is used for both $^{163}\mathrm{Ho}$ and $^{163}\mathrm{Dy}$ and the magnetron and axial frequencies are sufficiently equal.
Thus, all systematic uncertainties in the free-space cyclotron frequency measurement cancel out to a large extent in the determination of the frequency ratio $R$ and are smaller than $10^{-12}$.
Extended Data Table~\ref{tab:systematics} summarizes the considered systematic shifts.
An additional systematic uncertainty can arise from the fact that HCIs might have long-lived low-energy atomic metastable states, as observed in previous measurements~\cite{Schuessler20}. 
This is undesirable since it will shift the determined $Q$-value by the energy of the metastable state. 
In Section~\ref{sec:Qval} we compare the $Q$-values resulting from the measurements of three different charge states which allows us to exclude potential shifts of the $Q$-value due to long-lived electronic metastable states that would influence each charge state differently.
The final ratios of free cyclotron frequencies of the ions in the different charge states are summarized in Table~\ref{tab:results}.

\section{Calculation of binding energy differences}\label{sec:BindingEnergy}

Theoretical calculations provide the binding energies of the electrons removed from the neutral Ho and Dy atoms. 
The Dy and Ho atoms are in the [Xe]$4f^{10}6s^{2}$~${}^5 I_{8}$ and [Xe]$4f^{11}6s^{2}$~${}^4 I_{15/2}$ electronic states, respectively.
For a better control of systematic effects, several HCI of Dy and Ho were considered in the experiment, namely, Dy${}^{38+,39+,40+}$, with the ground states [Ar]$3d^{10,9,8}$, respectively, and Ho${}^{38+,39+,40+}$ with [Ar]$3d^{10}4s$, [Ar]$3d^{10}$, and [Ar]$3d^{9}$, respectively.

\paragraph{Configuration interaction method} In a first set of calculations, the binding energies are calculated in Quanty \cite{Haverkort12, Quanty, Brass20} using the configuration interaction (CI) method. 
The starting point is a fully relativistic density functional theory (DFT) calculation with the full-potential local-orbital minimum-basis code FPLO \cite{Koepernik99, Opahle99, Eschrig04}. 
The DFT calculation determines the ground-state density of the ion around which a CI expansion is made. 
The corresponding Kohn-Sham orbitals are used as single particle basis to construct the Slater determinants that span a configuration space. 
The Hamiltonian comprises Coulomb and static Breit interaction between the electrons as well as their relativistic kinetic energies and potential energies due to Coulomb attraction of the ion's nucleus. 
Diagonalization of this Hamiltonian on a given configuration space using the L\'anczos algorithm determines the ground-state energy of an ion. 
	
At first, only the space of the ground state configuration is considered. 
Subsequently the configuration space is iteratively expanded to include single, double and triple excitations of electrons into orbitals with higher principal quantum numbers. 
Details of these calculations are given in the Methods section. We arrive to the calculated binding energy differences given in Table~\ref{tab:bindingenergydiff}.

\paragraph{Multiconfiguration Dirac-Hartree-Fock method} 

In the second set of calculations, we use the multiconfiguration Dirac-Hartree-Fock method (MCDHF)~\cite{Grant1970} and its combination with Brillouin-Wigner many-body perturbation theory~\cite{Kotochigova2007,GRASP2018}.

In the MCDHF method, the atomic state function is modeled as a superposition of configuration state functions (CSFs) with fixed angular momentum, magnetic and parity quantum numbers.
The CSFs are built as Slater determinants of Dirac orbitals in the $jj$ coupling scheme. 
Using the parallel GRASP2018 codes~\cite{GRASP2018}, we expand the space of virtual orbitals used for the construction of CSFs by single and double electron exchanges in a systematic manner. 
The convergence of the energies with respect to the maximal principal quantum number of virtual orbitals is monitored, and the spread of values resulting from different correlation models is used as a measure of the leading contribution (90\%) of the theoretical uncertainties.
In case of the HCI, the set of CSFs is generated with exchanges including all occupied orbitals from 1$s$ on, and with virtual orbitals up to typically 10$h$. 
Virtual orbitals are optimized in a layer-by-layer fashion~\cite{Fischer2016,GRASP2018}.
Effects of the Breit interaction, recoil, and approximate quantum electrodynamic corrections are accounted for by the configuration interaction method using orbitals from the MCDHF procedure~\cite{GRASP2018}.
More details are given in the Methods.
We obtain the theoretical values of the binding energy differences listed in Table~\ref{tab:bindingenergydiff}.

In a third set of calculations, we use the Multiconfiguration Dirac-Fock General-Matrix-Elements (MCDFGME) code\cite{ild2005}, to check the previous results. 
The calculation is performed in the \emph{optimized level} mode, where all correlation  orbitals are fully relaxed instead of the layer by layer method. 
Convergence is much more difficult in this case and limits the number of extra orbitals that can be added in the evaluation of correlation.
In this calculation, the magnetic and retardation interaction at the Breit level are included in the Dirac-Fock equations on the same footing as the Coulomb interaction, meaning that the Breit interaction is included to all orders in the correlation energy \cite{ind1995}. 
The Uehling potential is also evaluated to all orders \cite{ind2013}.
Finally, self-energy screening is calculated  using both the Welton method \cite{igd1987} and the model operator method \cite{sty2013}.
For the HCIs, energies obtained by exciting occupied orbitals from $3s$ or $3d$ to open shells  ($4f$, $6p$, $5d$, $7s$, $7p$ and  $5g$) were compared. 
For neutrals, values obtained by exciting the core from $3d$ and $4s$ were compared.
Calculations included only single and double excitations, as triple excitations lead to unmanageably large numbers of magnetic and retardation integrals. 
All possible single excitations were included, even those obeying Brillouin's theorem \cite{ild2005}.
The results are given in Table~\ref{tab:bindingenergydiff} and are in good agreement with the GRASP2018 evaluation.
Both sets of values are in agreement with the uncorrelated values \cite{risp2004}, confirming the good compensation of correlation between the two ions.

\paragraph{Final values for the binding energy difference} 
The final binding energy $\Delta E_{\mathrm{B}}^{q+}$ for each charge state $q$ is calculated as the weighted average of the values from the CI and MCDHF calculations (cf.~Table~\ref{tab:bindingenergydiff}).
The uncertainty is determined by comparing the inner and outer errors and using the larger one as final uncertainty on $\Delta E_{\mathrm{B}}^{q+}$. 
For the charge states $q = \{39,\,40\}\cdot e$, the larger of the two uncertainties is averaged with the uncertainty assuming correlations between the CI and MCDHF methods, i.e. with the uncertainty of $0.8\,\mathrm{eV}$ of the MCDHF method.
The resulting $\Delta E_{\mathrm{B}}^{q+}$ are consistent with the MCDFGME calculations described above as well as with the calculations recently published in~\cite{Savelyev23}.

\begin{table*}[htbp]
	\centering
	\begin{tabular}{lrrrr}
		\hline
			$q/e$ & $\Delta E_{\mathrm{B, CI}}$ & $\Delta E_{\mathrm{B, MCDHF}}$ & $\Delta E_{\mathrm{B, MCDFGME}}$ & $\Delta E_{\mathrm{B}}^{q+}$ \\
		\hline
			$38$ & $38.8 \pm 1.0$ & $36.5 \pm 0.8$ & $38.1 \pm 1.5$ & $37.4 \pm 1.4$ \\
			$39$ & $1148.2 \pm 1.0$ & $1146.7 \pm 0.8$ & $1148.1 \pm 1.5$ & $1147.3 \pm 0.7$ \\
			$40$ & $1116.6 \pm 1.0$ & $1115.1 \pm 0.8$ & $1116.4 \pm 1.5$ & $1115.7 \pm 0.7$ \\
		\hline
	\end{tabular}
	\caption{Summary of the electronic binding energy differences for the three charge states (first column) in electron volts (eV) from the three theory calculations: Configuration interaction (CI) method (second column), the Multiconfiguration Dirac-Hartree-Fock (MCDHF) method (third column) and the Multiconfiguration Dirac-Fock General-Matrix-Elements (MCDFGME) method (fourth column).
	The given uncertainties correspond to the $1\sigma$ uncertainty.
	In the last column the final $\Delta E_{\mathrm{B}}$ for the determination of the $Q$-value are given which were calculated as the  weighted average of the CI and MCDHF methods.
	For details on the calculation of the uncertainties see main text.}
	\label{tab:bindingenergydiff}
\end{table*}

\section{$Q$-value determination}\label{sec:Qval}

The $Q$-value of the electron capture in $^{163}\mathrm{Ho}$ is determined from the measured ratio of the free cyclotron frequencies $R$ (see Section~\ref{sec:PTMS}) and the theoretically calculated binding energy differences $\Delta E_{\mathrm{B}}^{q+}$ (see Section~\ref{sec:BindingEnergy} and Table~\ref{tab:bindingenergydiff}) for each charge state $q = \{38, 39, 40\} \cdot e$ according to Equation~\ref{eq:Qvaluecalculation}.
The (reference) mass $m_{\mathrm{Dy}}^{q+}$ of $^{163}\mathrm{Dy}^{q+}$ is calculated starting from the mass of atomic $^{163}\mathrm{Dy}$, $m_{\mathrm{Dy}}$~\cite{Wang21}, and subtracting the masses of the $n$ missing electrons~\cite{Tiesinga21, Sturm14} and their binding energies~\cite{Kramida21}.
Table~\ref{tab:results} lists the ratios $R$ for the three measured charge states as well as the $1\sigma$ uncertainty $\delta R$ which is computed using standard Gaussian uncertainty propagation.

Using Equation~(\ref{eq:Qvaluecalculation}) and the binding energy differences, the final $Q$-values are calculated for the three charge states and are summarized in Table~\ref{tab:results}.

\begin{table*}[htbp]
	\centering
	\begin{tabular}{lrrrr}
		\hline
			$q/e$ & $R$ & $\delta R$ & $\Delta E_{\mathrm{B}}^{q+} (\mathrm{eV})$ &  $Q$ ($\mathrm{eV}$/$c^2$) \\
		\hline
			38 & 1.000000018623 & 3.0E-12 & $37.4 \pm 1.4$ & $2863.4 \pm 1.5$ \\
			39 & 1.000000011307 & 4.1E-12 & $1147.3 \pm 0.7$ & $2863.2 \pm 0.9$ \\
			40 & 1.000000011516 & 3.5E-12 & $1115.7 \pm 0.7$ & $2863.2 \pm 0.9$ \\
		\hline
	\end{tabular}
	\caption{Summary of the main results for the three charge states (column one): Weighted averages of the ratios (second column) and their uncertainty (third column), the weighted averages of the binding energy differences (fourth column, c.f. Table \ref{tab:bindingenergydiff}) and the calculated $Q$-values (fifth column) for the three measured charge states. 
	Uncertainties correspond to the $1\sigma$ statistical uncertainty.
	For details on the calculation of the uncertainties see main text.}
	\label{tab:results}
\end{table*}

The resulting $Q$-values for the different charge states agree within their $1\sigma$ uncertainties.
Resulting from the very good agreement, systematic deviations from either the free cyclotron ratio measurement or from the calculation of the binding energy difference can be excluded to a large extent. 
Furthermore, also the influence of unknown metastable electronic states can be largely ruled out since it is very unlikely that an electronic metastable state has exactly the same excitation energy in all three of the measured charge states.

The final $Q$-value is calculated as the weighted average of the $Q$-values obtained for the three charge states resulting in:

\begin{equation}
Q = 2863.2(0.6) \, \mathrm{eV}/c^2.
\end{equation}

\begin{figure}[htbp!]
	\centering
		\includegraphics[width=0.65\textwidth]{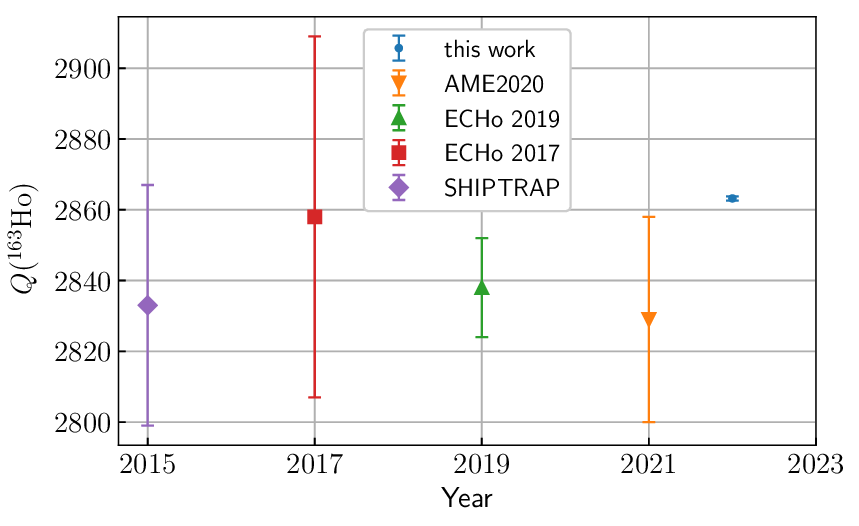}
	\caption{Comparison of the most recent measurements of the $^{163}\mathrm{Ho}$ EC $Q$-value from PTMS (SHIPTRAP~\cite{Eliseev15} and ``this work''), microcalorimetry (ECHo 2017~\cite{Ranitzsch17} and ECHo 2019~\cite{Velte19}) and the most recent AME adjustment 2020 \cite{Wang21}.}
	\label{fig:comparison}
\end{figure}

This value is in $1\sigma$ agreement with the previously measured value at SHIPTRAP of $2833(34)\mathrm{eV}/c^{2}$~\cite{Eliseev15} but 50 times more precise.
In Figure~\ref{fig:comparison} the most recent measurements of the $Q$-value of $^{163}\mathrm{Ho}$ from cryogenic microcalorimetry, PTMS and the Atomic Mass Evaluation (AME) 2020 are shown. 
The value from the AME 2020 is an average from three different microcalorimetric measurements.
Our value is slightly higher than the current AME adjustment and agrees within $1.2\sigma$.

The $^{163}\mathrm{Ho}$ $Q_{\mathrm{EC}}$-value was obtained by combining a high-precision measurement of the free-space cyclotron frequency of HCIs of the mother and daughter nuclide in a Penning trap and precise atomic physics calculations of the electronic binding energies of the missing electrons.
Experiments investigating the electron neutrino mass by microcalorimetric measurements of the decay spectrum of $^{163}\mathrm{Ho}$ such as those of the ECHo and HOLMES collaborations are now provided with an independently measured $Q$-value with an unprecedented precision of $0.6\,\mathrm{eV}$, which allows the assessment of systematic uncertainties in the neutrino mass determination using cryogenic microcalorimetry on a level of $<1\,\mathrm{eV}$.

\newpage
\section{Acknowledgements}

We acknowledge funding and support from the Max-Planck-Gesellschaft (C.S., V.D., M.D., P.F., Z.H., J.H., C.H.K., K.K., D.L., Y.N.N., A.R., R.X.S., S.E., K.B.)  and the International Max-Planck Research School for precision tests of fundamental symmetries (IMPRS-PTFS) (C.S., M.D., K.K.). 
This project was furthermore funded by the European Research Council (ERC) under the European Union’s Horizon 2020 Research and Innovation Programme under Grant Agreement No. 832848-FunI (K.K, K.B.) and 824109-EMP (C.E.), the Deutsche Forschungsgemeinschaft (DFG) - Project-ID 273811115-SFB1225-ISOQUANT (M.B., V.D., M.W.H., C.S.), the Deutsche Forschungsgemeinschaft through grant No. INST 40/575-1 FUGG (JUSTUS 2 cluster) (M.B., M.W.H.), the Deutsche Forschungsgemeinschaft Research Unit FOR2202 Neutrino Mass Determination by Electron Capture in 163Ho, ECHo (funding under Grant No. HA 6108/2-1 (M.B., M.W.H.), GA 2219/2-2 (L.G.), EN 299/7-2 (C.E.), EN299/8-2 (C.E.), BL 981/5-1 (K.B.), DU 1334/1-2 (C.E.D., H.D., D.R.)) , the Max-Planck-RIKEN-PTB Center for Time, Constants and Fundamental Symmetries (K.B.) and the state of Baden-Württemberg through bwHPC (M.B., M.W.H.). 
K.B., P.I. and Y.N.N. are members of the Allianz Program of the Helmholtz Association, contract number EMMI HA-216 ``Extremes of Density and Temperature: Cosmic Matter in the Laboratory''.

The results presented in this paper are based on work performed before Feb., 24th 2022.
This work comprises parts of the Ph.D. thesis work of C.S. to be submitted to Heidelberg University, Germany.

\newpage
\section{Author contributions statement}

\paragraph{Conceived and designed the experiments} Ch. Schweiger, M. Door, Ch. E. Düllmann, C. Enss, P. Filianin, L. Gastaldo, Yu. N. Novikov, A. Rischka, R.X. Schüssler, S. Eliseev, K. Blaum.

\paragraph{Performed the experiments} Ch. Schweiger, M. Door, P. Filianin, J. Herkenhoff, K. Kromer, S. Eliseev.

\paragraph{Analyzed the data} Ch. Schweiger, S. Eliseev.

\paragraph{Performed and evaluated the theoretical calculations} M. Braß, V. Debierre, Z. Harman, M.W. Haverkort, P. Indelicato

\paragraph{Supervision, interpretation and discussion of the theoretical calculations} M. Braß, V. Debierre, Z. Harman, M.W. Haverkort, C.H. Keitel, P. Indelicato

\paragraph{Contributed materials/analysis tools} Ch. Schweiger, M. Door, H. Dorrer, Ch. E. Düllmann, K. Kromer, D. Lange, D. Renisch, S. Eliseev, K. Blaum.

\paragraph{Wrote the paper} Ch. Schweiger, M. Braß, V. Debierre, Z. Harman, M.W. Haverkort, P. Indelicato, S. Eliseev.
\newline
\newline
\noindent
All authors took part in the critical review of the manuscript.

\newpage
\bibliography{bibliography}% common bib file
%% if required, the content of .bbl file can be included here once bbl is generated
%%\input HoDyPaperNaturePhysics.bbl

\section{Methods}\label{sec:methods}

\paragraph{Measurement preparation and sequence} 
The measurement preparation starts with loading a set of three ions, in the order $^{163}\mathrm{Dy}$, $^{163}\mathrm{Ho}$ and $^{163}\mathrm{Dy}$ into traps~2,~3, and~4, respectively (cf.~Fig.~\ref{fig:ExampleData}~(a)).
Each HCI is first loaded into trap 2 where its motional amplitudes are reduced by resistive cooling~\cite{Brown86, Cornell90}.
Great care is taken to ensure that only a single HCI is captured in a trap and cooled. 
For this also the ``magnetron cleaning'' technique is applied~\cite{Heisse19}.
After being prepared in this way, the ion is moved to one of the following traps and stored until the set of ions for a measurement is complete. 

In both measurement traps (traps~2 and~3), the motional frequencies of the HCIs are measured using the single-dip, double-dip and Pulse-and-Phase (PnP) techniques~\cite{Cornell89, Cornell90}.
The magnetron frequency is small compared to the other two motional frequencies and depends only very weakly on the ion's mass and is therefore measured only once a day using the double-dip technique prior to the main measurement sequence.
Thus, the main measurement sequence reduces to a measurement of the modified cyclotron frequency (PnP technique) and the axial frequency (double-dip technique) which are performed simultaneously in traps 2 and 3.
Compared to a single-trap measurement, this effectively doubles the statistics and furthermore allows different analysis methods to be employed as well as systematic checks by comparing the results obtained in both traps.

In the PnP cycle, the starting phase of the cyclotron motion is set by exciting it using a dipolar pulse with the frequency determined with the double-dip method during the preparation.
The modified cyclotron motion then evolves freely during the phase evolution time $T_{\mathrm{evol}}$ (about $40\,\mathrm{s}$) while the axial frequency is determined using a dip measurement~\cite{Rischka20, Filianin21, Kromer22}.
Following the phase evolution time, the phase information that accumulated in the modified cyclotron motion is coupled to the axial motion using a $\pi$-pulse on the sideband frequency and the final phase is measured with the image current detection system~\cite{Cornell89}.
This is done in traps 2 and 3, starting with the ions in position 1.
Subsequently, the ions are shuttled to position 2, which effectively swaps the ion species in traps 2 and 3 and the measurement is repeated (cf.~Fig.~\ref{fig:ExampleData}~(a)).
This sequence is repeated 24 times in one main measurement loop and can be continued in principle infinitely long.
Typically, the measurement is stopped due to either external magnetic field perturbations or charge exchange of the HCIs.
Lifetimes of the HCIs until a charge exchange process happens are up to 36 hours.
Reloading ions is beneficial since it allows one to compare different sets of ions and therefore also systematic checks for contaminant ions that might be present in the Penning traps during the measurement or for possible metastable electronically excited states in the HCIs~\cite{Schuessler20}.

\paragraph{Convergence studies with the Configuration Interaction and Multiconfiguration Dirac-Dock methods}
In the configuration interaction calculations with the Quanty code, we iteratively expanded the configuration space with single, double and triple excitations into single-electron states with higher principal quantum numbers. 
Explicitly for the ions this implies iterative inclusion of excitations into orbitals with $n=4,5,6,7,8$ and for the neutral atom $n=5,6$. 
	
The evolution of ground-state energy with expanding configuration space is monitored and shows to good approximation a $1/n$ behaviour which allows extrapolation of the ground-state energy to estimate its uncertainty due to a truncated configuration space. 
Considering further uncertainties due to numerical accuracy, choice of single particle basis sets and triple excitations, we arrive at a total uncertainty of $1~\mathrm{eV}$ for the estimation of the differences in binding energies of Ho$^{q+}$ and Dy$^{q+}$. 
	
As consistency check, for every step where the configuration space is increased the binding energy difference between Ho$^{q+}$ and Dy$^{q+}$ is calculated. 
It shows an approximate $1/n^2$ behavior, which again allows for extrapolation. 
Within our uncertainties we obtain the same results as in Table~\ref{tab:bindingenergydiff} in the article.

In case of the Multiconfiguration Dirac-Fock calculations with the GRASP2018 package, as described in the article, we have found that for neutral atoms, the inclusion of all spectroscopic orbitals into the active space would lead to several tens of millions of CSFs and is currently not tractable. 
For these systems, we include exchanges from the $3s$ orbital up to typically $8h$.
To bridge the different models used for the neutrals and the HCIs, we also study the intermediate Pd-like HCIs Dy${}^{20+}$ and Ho${}^{21+}$, with excitations from the $2s$ orbital to typically $10h$.
We observe that correlation terms largely cancel in energy differences such as $[E(\textrm{Ho})-E(\textrm{Ho}^{21+})] - [E(\textrm{Dy})-E(\textrm{Dy}^{20+})]$ and $[E(\textrm{Ho}^{21+})-E(\textrm{Ho}^{40+})] - [E(\textrm{Dy}^{20+})-E(\textrm{Dy}^{40+})]$ due to structural similarities of nearby charge states.
These differences converge more quickly when extending the set of virtual orbitals than the individual energies $E(\textrm{Ho})$ and $E(\textrm{Dy})$ of the neutrals.
We note that such an inclusion of an intermediate system is useful because of the high charge states $38+$, $39+$ and $40+$ in the experiment, and allows reducing uncertainties.

\section{Extended Data}

\begin{table}[htbp]
\centering
\begin{tabular}{ll}  
\toprule
Systematic shift 										& Magnitude \\
\midrule
Relativistic shift 									& $\delta R < 10^{-12}$~\cite{Ketter14a} \\
Field anharmonicities/imperfections & $\delta R < 10^{-13}$~\cite{Roux12, Ketter14b} \\
Image charge shift						      & $\delta R < 10^{-13}$~\cite{Roux12, Schuh19} \\
Dip lineshape									      & $\delta R < 10^{-13}$ \\
$C_{1}B_{1}$     									  & Effect cancels in the ratio. \\
$C_{1}C_{3}$                        & Effect cancels in the ratio. \\
\bottomrule
\end{tabular}
\caption{Overview of the considered systematic shifts in the determination of the frequency ratio.
The relativistic shift was estimated assuming conservatively that both radii agree within 1\%.}
\label{tab:systematics}
\end{table}

\end{document}